\RequirePackage{fix-cm}
\documentclass[natbib,smallextended]{svjour3}
\smartqed
\usepackage{natbib}
\usepackage{mathptmx}
\usepackage{aps-bibstyle}
\usepackage{amssymb}
\usepackage{graphicx}
\usepackage{mathptmx}
\usepackage{aas_macros}
\usepackage{epstopdf}
\usepackage{ccicons}

\journalname{\ssr}

\begin{document}

\newcommand{\dtsifac}{{\rm \Delta{}TSI_{fac}}}
\newcommand{\dtsispt}{{\rm \Delta{}TSI_{spt}}}
\newcommand{\wms}{{\rm Wm^{-2}}}

\title{Solar cycle variation in solar irradiance}
\author{K.~L. Yeo \and N.~A. Krivova \and S.~K. Solanki}
\institute{
K.~L. Yeo \and N.~A. Krivova \and S.~K. Solanki \at Max-Planck-Institut f\"{u}r Sonnensystemforschung, 37077 G\"{o}ttingen, Germany \\ \email{yeo@mps.mpg.de}
\and
S.~K. Solanki \at School of Space Research, Kyung Hee University, Yongin, 446-701 Gyeonggi, Korea 
}
\date{Received: 14 March 2014 / Accepted: 17 June 2014}
\maketitle

\begin{abstract}
The correlation between solar irradiance and the 11-year solar activity cycle is evident in the body of measurements made from space, which extend over the past four decades. Models relating variation in solar irradiance to photospheric magnetism have made significant progress in explaining most of the apparent trends in these observations. There are, however, persistent discrepancies between different measurements and models in terms of the absolute radiometry, secular variation and the spectral dependence of the solar cycle variability. We present an overview of solar irradiance measurements and models, and discuss the key challenges in reconciling the divergence between the two.
\end{abstract}

\section{Introduction}

The 11-year solar activity cycle, the observational manifest of the solar dynamo, is apparent in indices of solar surface magnetism such as the sunspot area and number, 10.7 cm radio flux and in the topic of this paper, solar irradiance. The observational and modelling aspects of the solar cycle are reviewed in \cite{hathaway10} and \cite{charbonneau10}, respectively. Solar irradiance is described in terms of what is referred to as the total and spectral solar irradiance, TSI and SSI. They are defined, respectively as the aggregate and spectrally resolved solar radiative flux (i.e., power per unit area and power per unit area and wavelength) above the Earth's atmosphere and normalized to one AU from the Sun. By factoring out the Earth's atmosphere and the variation in the Earth-Sun distance, TSI and SSI characterize the radiant behaviour of the Earth-facing hemisphere of the Sun.

The variation of the radiative output of the Sun with solar activity has long been suspected \citep{abbot23,smith75,eddy76}. However, it was not observed directly till satellite measurements, free from atmospheric fluctuations, became available. TSI and SSI, at least in the ultraviolet, have been monitored regularly from space through a succession of satellite missions, starting with Nimbus-7 in 1978 \citep{hickey80,willson88,frohlich06,deland08,kopp12}. A connection between variations in TSI and the passage of active regions across the solar disc was soon apparent \citep{willson81,hudson82,oster82,foukal86}, leading to the development of models relating the variation in solar irradiance to the occurrence of bright and dark magnetic structures on the solar surface. While not the only mechanism mooted, models that ascribe variation in solar irradiance at timescales greater than a day to solar surface magnetism have been particularly successful in reproducing observations \citep{domingo09}. At timescales shorter than a day, excluded from this discussion, intensity fluctuations from acoustic oscillations, convection and flares begin to dominate \citep{hudson88,woods06,seleznyov11}.

The measurement and modeling of the variation in solar irradiance over solar cycle timescales, a minute proportion of the overall level (about $0.1\%$ in the case of TSI), is a substantial achievement. Though significant progress has been made over the past four decades, considerable discrepancies remain between different measurements and models in terms of the absolute radiometry, secular variation and the spectral dependence of the cyclical variability \citep[see the recent reviews by][]{ermolli13,solanki13}. In the following, we present a brief overview of the current state of solar irradiance observations (Sect. \ref{measurements}) and models (Sect. \ref{models}). We then discuss the key issues in reconciling measurements and models (Sect. \ref{discussionssr}) before giving a summary (Sect. \ref{summaryssr}). Our focus will be on the far-ultraviolet to the infrared region of the solar spectrum. This is the spectral range where the bulk of the energy in solar radiation is confined and commonly covered by present-day models aimed at returning both TSI and SSI. We refer the reader to \cite{lilensten08,woods08} for an overview of solar cycle variability in the extreme-ultraviolet and shortwards, and \cite{tapping13,dudokdewit14} for radio wavelengths.

\section{Measurements}
\label{measurements}

\subsection{Total solar irradiance, TSI}
\label{tsimeasurements}

\begin{figure}
\includegraphics[width=\textwidth,trim=0cm .4cm 0cm 0cm,clip=true]{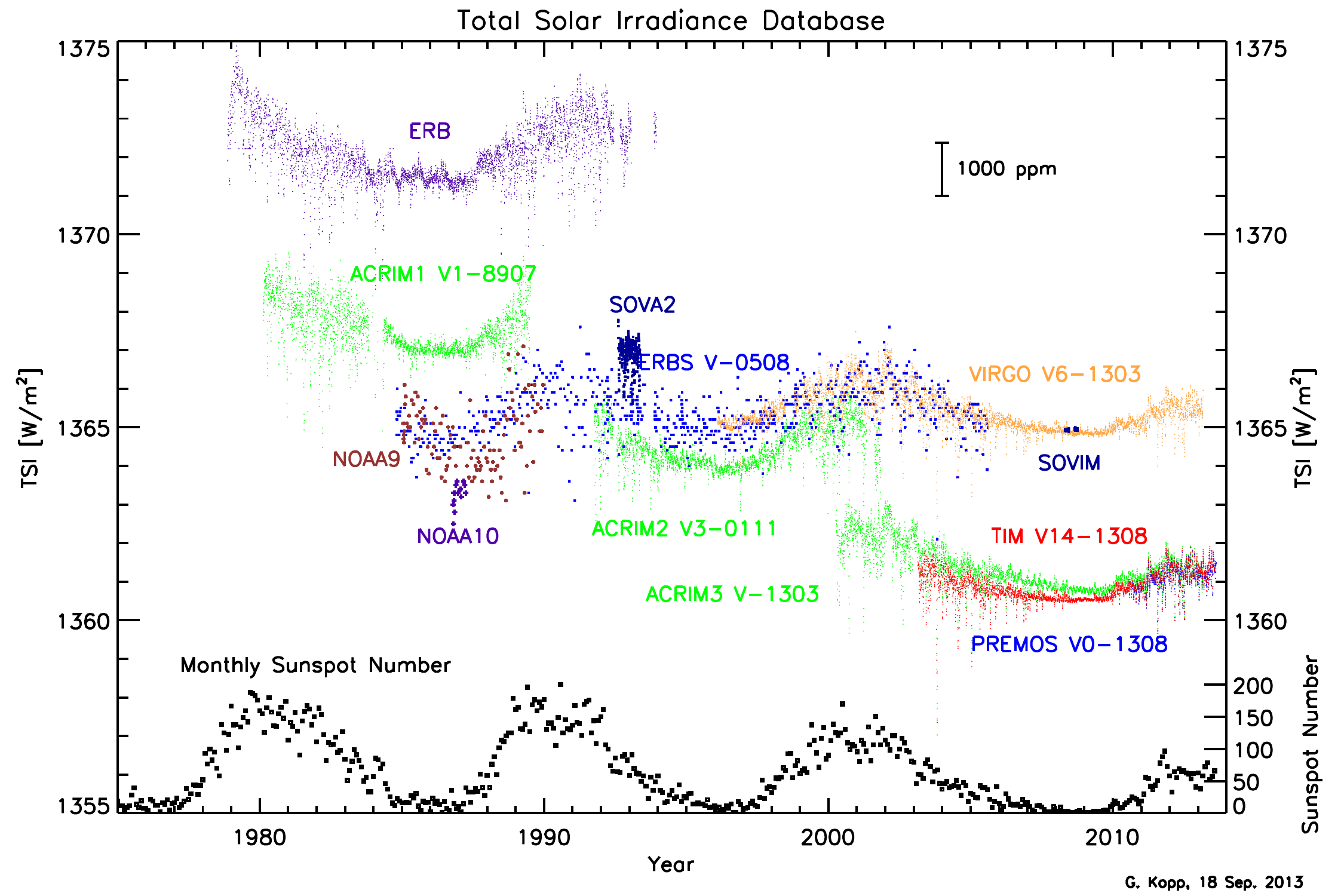}
\caption{The measurements from the succession of TSI radiometers sent into orbit since 1978 (colour coded) and the monthly mean of the sunspot number (black, lower right axis). Each TSI record is annotated by the name of the instrument and where applicable, the version number and/or the year and month of the version. Courtesy of G. Kopp (http://spot.colorado.edu/~koppg/TSI/).}
\label{tsi}
\end{figure}

The measurements from the succession of TSI radiometers sent into space since 1978, collectively representing a nearly uninterrupted record, exhibit clear solar cycle modulation. This is illustrated in Fig. \ref{tsi} by the comparison between the various TSI records and the monthly sunspot number. All these instruments are based on active cavity electrical substitution radiometry \citep{butler08,frohlich10}. Succinctly, TSI is measured by allowing solar radiation into a heated absorptive cavity intermittently and adjusting the heating power as necessary to maintain thermal equilibrium. While these observations are sufficiently stable over time to trace solar cycle variability, only about $0.1\%$ of the overall level, the measurements from the various instruments are offset from one another by a greater margin, reflecting the uncertainty in the absolute radiometry.

With the early instruments, specifically Nimbus-7/ERB \citep{hickey80,hoyt92}, SMM/ACRIM1 \citep{willson79} and ERBS/ERBE\footnote{Nimbus-7/ERB denotes the Earth Radiation Budget instrument onboard Nimbus-7, SMM/ACRIM1 the Active Cavity Radiometer Irradiance Monitor onboard the Solar Maximum Mission and ERBS/ERBE the Earth Radiation Budget Experiment onboard the similarly named satellite.} \citep{lee87}, the spread in absolute radiometry arose mainly from the uncertainty in the aperture area \citep{frohlich12,ermolli13}. As the determination of the aperture area improved, so the absolute radiometry from the succeeding missions converged. That is, up till the Total Irradiance Monitor, TIM \citep{kopp05a,kopp05b,kopp05c} onboard the SOlar Radiation and Climate Experiment, SORCE, launched in 2003 \citep{rottman05}.

The measurements from TIM were about $5\:\wms$ lower than the concurrent observations from ACRIMSAT/ACRIM3\footnote{The ACRIM radiometer onboard the ACRIM SATellite.} \citep{willson03} and SoHO/VIRGO\footnote{The Variability of IRradiance and Gravity Oscillations experiment onboard the Solar and Heliospheric Observatory.} \citep{frohlich95,frohlich97}. Tests conducted at the TSI Radiometer Facility, TRF \citep{kopp07} with ground copies of ACRIM3, TIM and VIRGO revealed unaccounted stray-light effects in ACRIM3 and VIRGO \citep{kopp11,kopp12,fehlmann12}. Correction for scattered light subsequently introduced to the ACRIM3 record based on the results of these tests brought it down to within $0.5\:\wms$ of the TIM record. The similar proximity between the measurements from the PREcision MOnitor Sensor, PREMOS onboard Picard, launched in 2010 \citep{schmutz09,schmutz09b,fehlmann12}, with TIM radiometry provided further evidence that the lower absolute level first registered by TIM is likely the more accurate. In a first, PREMOS was calibrated in vacuum at full solar power levels prior to flight at two separate facilities, the National Physical Laboratory and the TRF. As such, it is considered to be more reliably calibrated than preceding TSI radiometers.

\begin{figure}
\includegraphics[width=\textwidth]{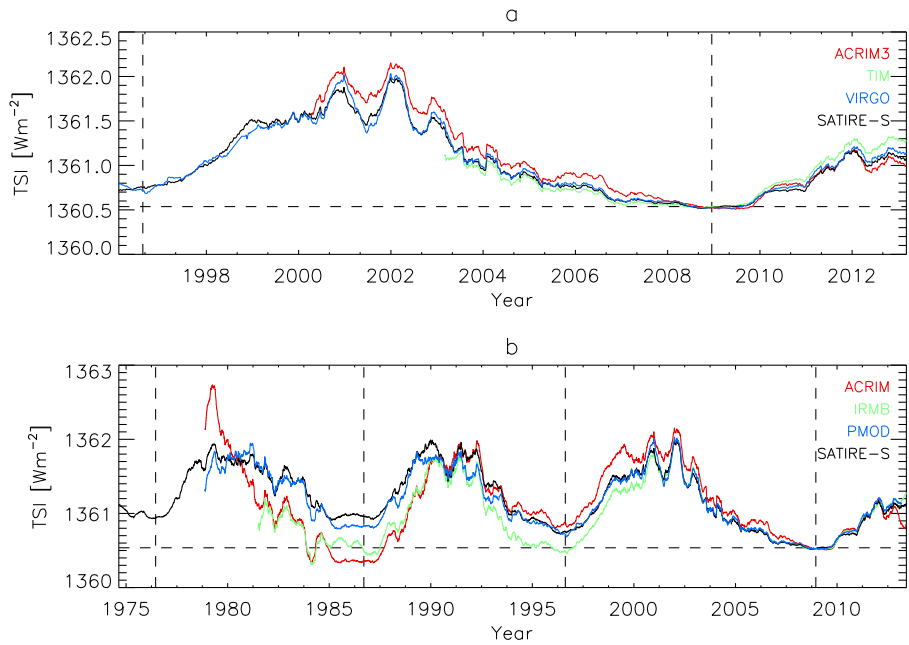}
\caption{a) TSI measurements from ACRIMSAT/ACRIM3 (version 11/13), SORCE/TIM (level 3, version 14) and SoHO/VIRGO (level 2, version $6\_002\_1302$). b) The ACRIM (version 11/13), IRMB (version dated 19th December 2013, provided by S. Dewitte) and PMOD (version ${\rm d}41\_62\_1302$) composite records of TSI. Also plotted is the SATIRE-S \citep{yeo14b} reconstruction of TSI. The vertical dashed lines mark the position of solar cycle minima. All the time series were normalized to TIM at the 2008 solar cycle minimum (horizontal dashed line) and smoothed with a 181-day boxcar filter.}
\label{comparetsi}
\end{figure}

Due to the limited lifetime of TSI radiometers, there is no single mission that covered the entire period of observation. Apart from ERBS (1984 to 2003) and VIRGO (1996 to present), there are no records that encompass a complete solar cycle minimum-to-minimum. Combining the measurements from the various missions into a single time series, obviously essential, is non-trivial due to ageing/exposure degradation, calibration uncertainty and other instrumental issues. The ACRIM, PREMOS, TIM and VIRGO instruments are designed with redundant cavities to allow in-flight degradation tracking. Even with this capability and the best efforts of the respective instrument teams, significant uncertainty persists over the long-term stability: conservatively, about $0.2\:\wms$ or 20 ppm per year \citep{solanki13}. This is visibly apparent in the discrepant amplitude of solar cycle variation in the ACRIM3, TIM and VIRGO records (Fig. \ref{comparetsi}a). Accounting for changes in instrument sensitivity, which are often particularly severe during early operation and can see discrete shifts such as that suffered by ERB and VIRGO, has proven to be particularly precarious \citep{hoyt92,lee95,dewitte04a,frohlich06}. It is worth mentioning here that the observations from the various radiometers do largely agree at solar rotational timescales, where apparent variability is much less affected by the instrumental influences discussed above.

There are, at present, three composite records of TSI, published by the ACRIM science team \citep{willson03}, IRMB\footnote{IRMB is the francophone acronym of the Royal Meteorological Institute of Belgium. This composite is also variously referred to as the RMIB or SARR composite.} \citep{dewitte04b,mekaoui08} and PMOD/WRC\footnote{The Physikalisch-Meteorologisches Observatorium Davos/World Radiation Center.} \citep{frohlich00,frohlich06}. These composites differ in terms of the amplitude of solar cycle variation, most readily apparent in the conflicting secular trend of the solar cycle minima level \citep[see][and Fig. \ref{comparetsi}b]{frohlich06,frohlich12}. The TSI reconstructions presented by \cite{wenzler09,ball12,yeo14b} based on the SATIRE-S model \citep{fligge00,krivova03,krivova11a} replicated, of these composites, the solar cycle and secular variation in the PMOD composite best (see Sect. \ref{modelsatires}).

\subsection{Spectral solar irradiance, SSI}
\label{ssimeasurements}

\subsubsection{Ultraviolet solar irradiance}
\label{ssimeasurements1}

\begin{figure}
\includegraphics[width=\textwidth,natwidth=3408,natheight=1388]{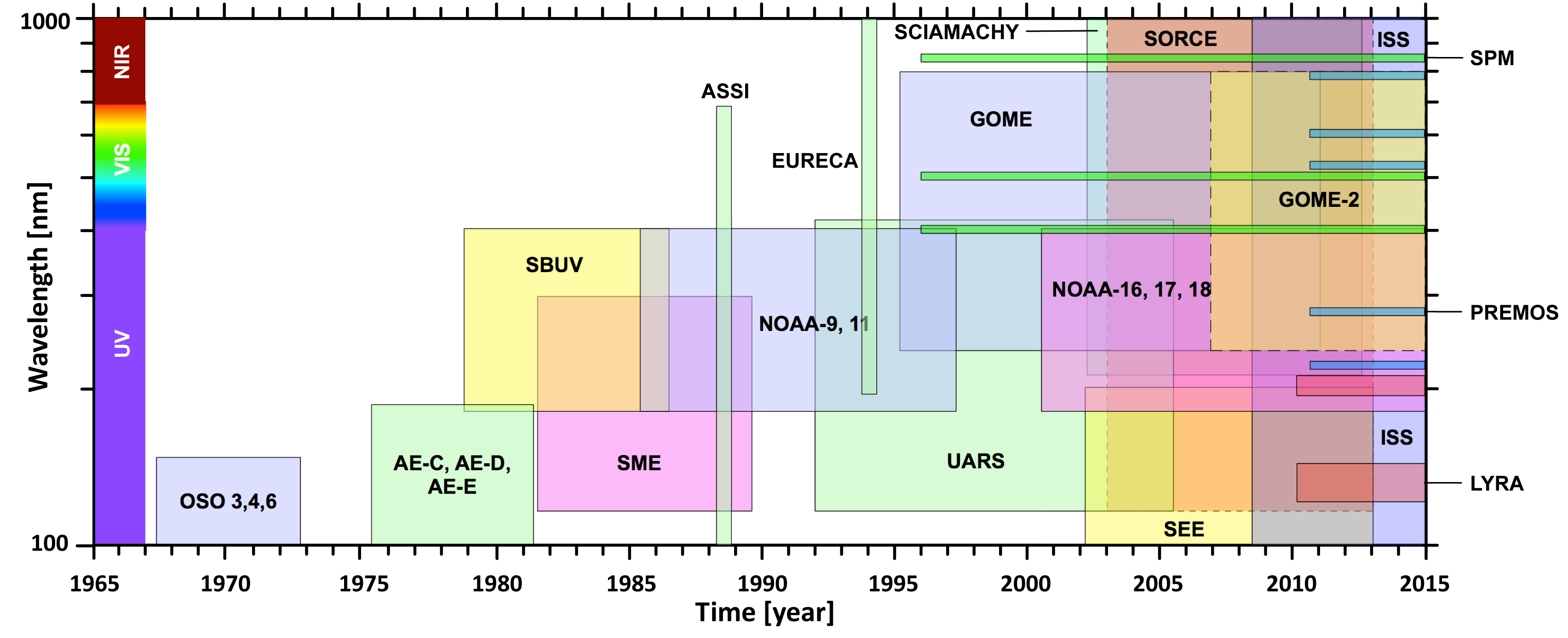}
\caption{Timeline and spectral range of the space missions making observations of the solar spectrum above 100 nm. Reproduced from \cite{ermolli13}, Creative Commons Attribution 3.0 License.}
\label{f03}
\end{figure}

The solar spectrum has been probed through a miscellany of spaceborne instruments over the past five decades (Fig. \ref{f03}) which differ in the regularity of measurements and the spectral range surveyed. As with TSI, SSI between around 120 and 400 nm has been monitored almost without interruption from space since 1978.

\begin{figure}
\includegraphics[width=\textwidth]{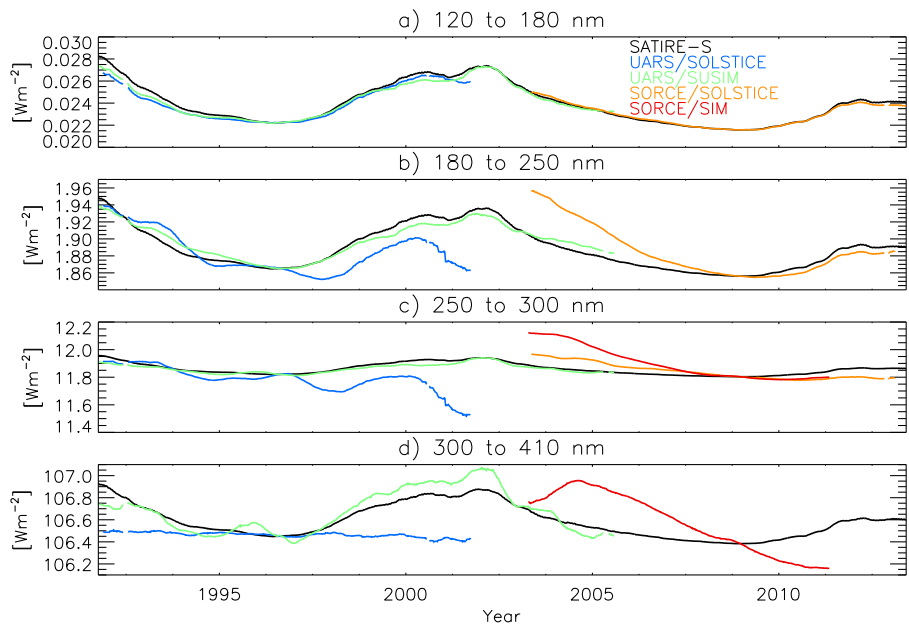}
\caption{Integrated solar irradiance in the annotated wavelength intervals in the SATIRE-S reconstruction of SSI (black). Also drawn are the measurements from the UARS and SORCE missions, rescaled to the SATIRE-S reconstruction at the 1996 and 2008 solar cycle minima, respectively. All the time series were smoothed by taking the 181-day moving average. Adapted from \cite{yeo14b}.}
\label{satireuarssorcessr}
\end{figure}

The key features of the body of ultraviolet solar irradiance measurements are similar to that of TSI, just discussed in Sect. \ref{tsimeasurements}. The observations from the various spectrometers display similar rotational variability but diverge in terms of the absolute radiometry and the amplitude of solar cycle variation, especially at wavelengths above 240 nm \citep[see Fig. \ref{satireuarssorcessr} and][]{deland12,unruh12,ermolli13,yeo14b}. In particular, the solar cycle amplitude in the measurements from the SIM \citep{harder05b,harder05a} and SOLSTICE \citep{mcclintock05,snow05a} experiments onboard SORCE is stronger than that registered by the SOLSTICE \citep{rottman01} and SUSIM \citep{brueckner93,floyd03} instruments onboard the predecessor mission, UARS\footnote{SIM denotes the Spectral Irradiance Monitor, SOLSTICE the SOLar STellar Irradiance Comparison Experiment, SUSIM the Solar Ultraviolet Spectral Irradiance Monitor and UARS the Upper Atmosphere Research Satellite.} by a factor of three to ten, depending on wavelength \citep[see Fig. 7 in][]{deland12}. This disparity, while broadly within the long-term uncertainty of said instruments, is much greater than encountered between pre-SORCE instruments \citep{deland12,ermolli13}. The long-term uncertainty of available ultraviolet solar irradiance observations is of similar magnitude as the variation over the solar cycle: on the order of 0.1 to $1\%$ per year, varying with wavelength and between instruments \citep{snow05a,merkel11,deland12}. The long-term uncertainty is also grossly greater than that afflicting TSI measurements. Again, due to the limited lifetime of spaceborne instrumentation, there is no record that extends beyond a complete solar cycle minimum-to-minimum with the exception of the observations from NOAA-9 SBUV/2\footnote{The second generation Solar Backscatter UltraViolet spectrometer onboard the ninth National Oceanic and Atmospheric Administration satellite.} \citep{deland98}.

The solar cycle modulation in and disparity between the various pre-SORCE records is illustrated in the time series plot by \citealt{deland08} (Fig. 2 in their paper), which is qualitatively analogous to Fig. \ref{tsi}. The authors presented the first published effort to compose the ultraviolet solar irradiance observations from these instruments into a single time series. The result, spanning 1978 to 2005, still contains overt instrumental trends for which the appropriate correction is not known. The challenge in the account of instrumental influences is substantially greater than with TSI, exacerbated by the wavelength dependence of instrumental effects and differences in the design, operation and calibration approach.

\subsubsection{SORCE/SIM SSI}
\label{ssimeasurements2}

The series of GOME instruments, the first of which was launched onboard ERS-2 in 1996 \citep{weber98,munro06} and ENVISAT/SCIAMACHY\footnote{GOME is short for the Global Ozone Monitoring Experiment, ERS-2 the second European Remote Sensing satellite and ENVISAT/SCIAMACHY the SCanning Imaging Absorption spectroMeter for Atmospheric CHartographY onboard the ENVIronmental SATellite.}, launched in 2002 \citep{skupin05a} made regular measurements of the solar spectrum in the 240 to 790 nm and 240 to 2380 nm wavelength range, respectively. These instruments are designed primarily for atmospheric sounding measurements which do not require absolute radiometry rather than solar irradiance monitoring. As such, they lack the capability to track instrument degradation in-flight, rendering the observations unsuitable for tracing the solar cycle variation of the solar spectrum reliably. The long-term stability of the narrowband (FWHM of 5 nm) photometry at 402, 500 and 862 nm from the Sun PhotoMeter, SPM on SoHO/VIRGO \citep{frohlich95,frohlich97} is similarly problematic, though considerable progress has been made \citep{wehrli13}.

\begin{figure}
\includegraphics[width=\textwidth]{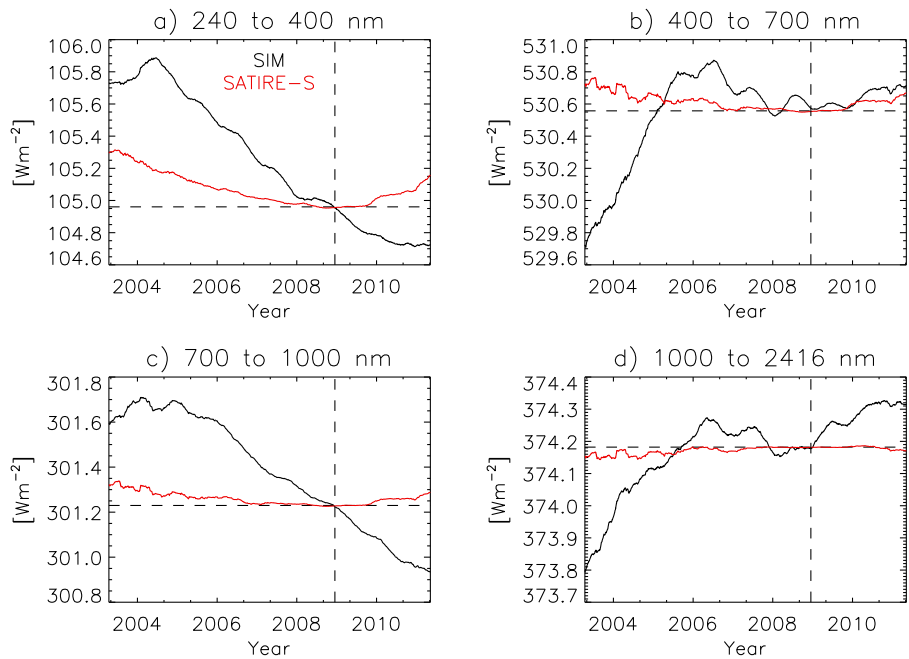}
\caption{181-day moving average of the integrated solar irradiance over the spectral intervals specified above each panel in the SORCE/SIM record (red) and in the SATIRE-S reconstruction (black). The SATIRE-S time series were rescaled to the corresponding SIM time series at the 2008 solar cycle minimum, the position and level at which is indicated by the dashed lines. Adapted from \cite{yeo14b}.}
\label{comparesatsimssr}
\end{figure}

\begin{figure}
\includegraphics[width=\textwidth]{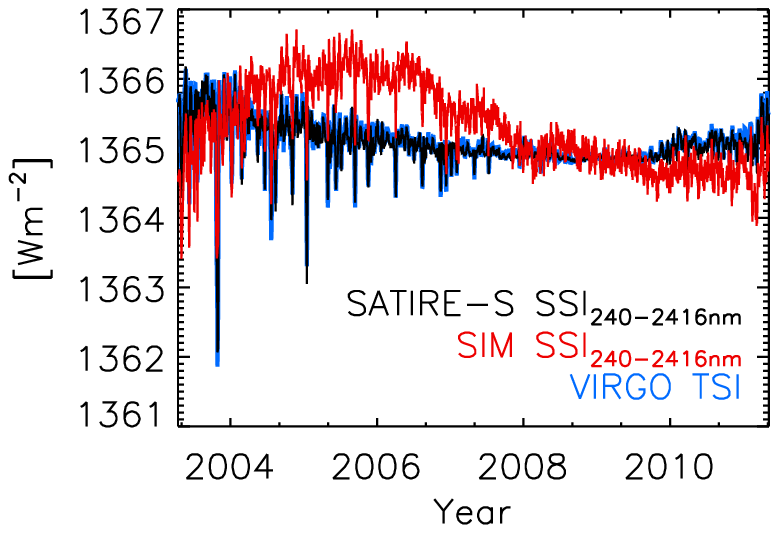}
\caption{Total flux registered by SIM (red), the integrated flux in the SATIRE-S reconstruction over a similar wavelength range (black) and VIRGO TSI (blue). The SIM and SATIRE-S time series were normalized to the VIRGO time series at the 2008 solar cycle minimum. Adapted from \cite{yeo14b}.}
\label{comparesatsimssr2}
\end{figure}

With these caveats in mind, one can say that regular monitoring of the solar spectrum longwards of the ultraviolet only started, in effect, with SIM. The instrument has been surveying the wavelength range of 200 to 2416 nm since 2003, providing the only continuous and extended record of the solar spectrum spanning the ultraviolet to the infrared presently available. The latest release (version 19, dated 23rd November 2013), covering the wavelength range of 240 to 2416 nm\footnote{SIM measurement from between 200 and 240 nm are not made publicly available in consideration of the fact that below about 260 nm, SIM observations start to register a lower signal-to-noise ratio than the concurrent observations from the SOLSTICE experiment onboard the same mission (J. Harder, private communication).}, is depicted in Figs. \ref{comparesatsimssr} and \ref{comparesatsimssr2}.

\begin{figure}
\includegraphics[width=\textwidth]{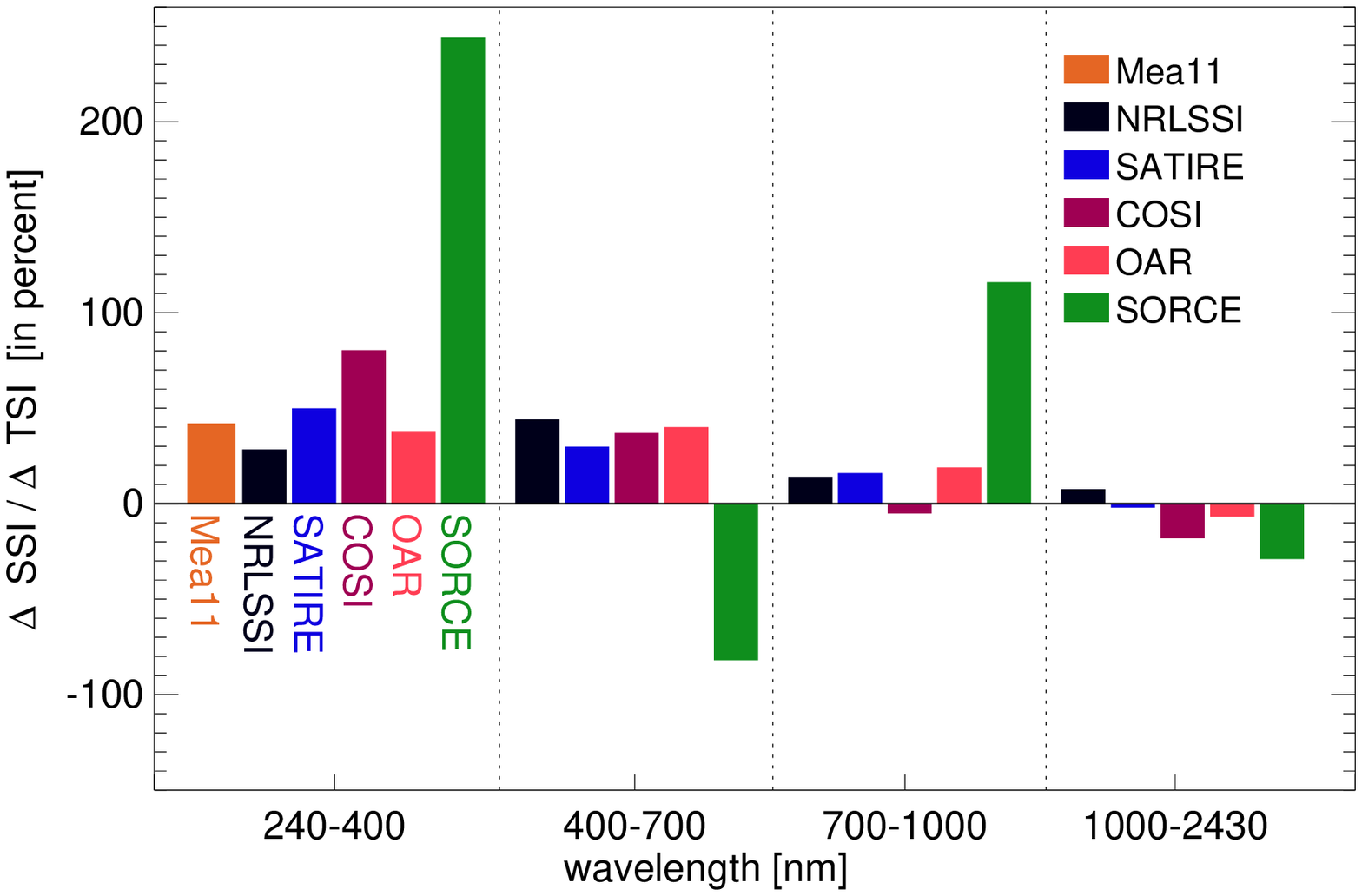}
\caption{Ratio of the variation in SSI over the same spectral intervals examined in Fig. \ref{comparesatsimssr}, ${\rm \Delta{}SSI}$, and the corresponding variation in TSI, ${\rm \Delta{}TSI}$, between solar cycle maximum and minimum (2002 and 2008) in the models discussed in Sect. \ref{models3}. For the model by \cite{morrill11}, denoted Mea11, we took ${\rm \Delta{}TSI}$ from the SATIRE-S model. Also depicted is the same for SORCE measurements between 2004 and 2008, with ${\rm \Delta{}SSI}$ and ${\rm \Delta{}TSI}$ from the SIM and TIM records, respectively. Adapted from \cite{ermolli13}, Creative Commons Attribution 3.0 License.}
\label{barchart}
\end{figure}

SIM returned rather unexpected results, precipitating the ensuing debate on the overall trend registered by the instrument \citep{harder09,ball11,deland12,unruh12,ermolli13,yeo14b}. Between 2004 and 2008, which is within the declining phase of solar cycle 23, SIM recorded a drop in ultraviolet flux (240 to 400nm, Fig. \ref{comparesatsimssr}a) that is almost double the decrease in TSI over the same period and multiple times greater than projections from pre-SORCE SSI measurements and models of solar irradiance (Fig. \ref{barchart}). Up to 2006, this pronounced downward trend in the ultraviolet is accompanied by a comparable increase in the visible (400 to 700 nm, Fig. \ref{comparesatsimssr}b), in anti-phase with the solar cycle. This is in conflict with SPM photometry (see next paragraph) and present-day models, all of which \citep[apart from][see Sect. \ref{modelsrpm}]{fontenla11} point to visible solar irradiance varying in phase with the solar cycle (Fig. \ref{barchart}). The variation in the infrared (700 to 2416 nm, Figs. \ref{comparesatsimssr}c and \ref{comparesatsimssr}d) between 2004 and 2008 is also significantly stronger than in model reconstructions (Fig. \ref{barchart}). The investigations of \cite{unruh08,unruh12,deland12,lean12,yeo14b} did, however, note that the rotational variability in SIM SSI is similar to that in pre-SORCE measurements and models.

Looking now at the full length of the SIM record, there is no constancy in how the overall trend at a given wavelength relate to the solar cycle (Fig. \ref{comparesatsimssr}). Measured solar irradiance is neither in phase nor anti-phase with the solar cycle. Apart from the solar cycle modulation evident in pre-SORCE ultraviolet solar irradiance measurements, this also runs counter to the positive correlation between SPM visible (500 nm) photometry and TSI over the similar period of 2002 to 2012 reported by \cite{wehrli13}. There being no other extended record of SSI covering a similar spectral range, it cannot be ruled out completely that segments of the solar spectrum may vary in a non-cyclic manner as apparent in SIM SSI. It is, however, almost irrefutable that the integral of the solar spectrum over all wavelengths, TSI, does exhibit solar cycle variation. As the spectral range surveyed by SIM accounts for more than $97\%$ of the power in solar radiation, the total flux recorded by the instrument should already replicate most of the variability in TSI but that is evidently not the case (Fig. \ref{comparesatsimssr2}). In contrast, the reconstruction of the integrated solar irradiance over the spectral range of SIM from SSI models replicates most of the variability in TSI \citep[see Fig. \ref{comparesatsimssr2} and][]{ball11,lean12,yeo14b}.

The discrepancies between SIM SSI and other measurements and models reported in various studies, summarized above, were taken to indicate that there are unaccounted instrumental trends in the SIM record \citep{ball11,deland12,lean12,unruh12,yeo14b}. This is favoured here over alternative interpretations such as the apparent trends between 2004 and 2008 implying a change in the physics of the Sun during this period compared to earlier times or that there are gaping insufficiencies in our understanding of the physical processes driving variations in solar irradiance. As we will argue in Sect. \ref{models1}, the decline in visible flux between 2003 and 2006 (Fig. \ref{comparesatsimssr}b) is not consistent with our current understanding of solar surface magnetism and its effect on solar irradiance.

Evidently, the direct observation of the variation in SSI over solar cycle timescales is afflicted by considerable uncertainty. The situation is set to improve with the continuing efforts to calibrate SCIAMACHY and SORCE spectrometry, and spectral measurements expected from ISS/SOLSPEC \citep[][]{thuillier09,thuillier14} and the upcoming JPSS\footnote{ISS/SOLSPEC denotes the SOLar SPECtrum experiment onboard the International Space Station and JPSS the Joint Polar Satellite System.} mission \citep{richard11}. JPSS, set to be launched in 2017, will carry an improved version of the SIM instrument. For more extensive reviews of the measurement of solar irradiance, we refer the reader to \cite{deland08,deland12,domingo09,kopp12,frohlich12,ermolli13,solanki13}.

\section{Models}
\label{models}

The body of satellite measurements of solar irradiance, while evidently core to our understanding of the solar cycle variation in the radiative output of the Sun, cover a limited period in time and suffer significant uncertainty. Models of solar irradiance serve both to complement these observations and to advance our understanding of the physical processes driving the apparent variability.

\subsection{Solar surface magnetism}
\label{models1}

A feature of the 11-year solar activity cycle is the cyclical emergence and evolution of kilogauss-strength magnetic concentrations in the photosphere \citep{solanki06}. The main properties of these magnetic concentrations can be explained by describing them as magnetic flux tubes \citep{spruit76,spruit83,solanki93a}. They range in physical extent from on the order of $10^{1}$ to $10^{5}$ km in cross section. The lower end corresponds to the small-scale magnetic elements which make up active region faculae and quiet Sun network/internetwork \citep{lagg10,riethmuller14}, and the upper end to sunspots and pores.

Solar irradiance is modulated by photospheric magnetic activity from its effect on the thermal/radiant property of the solar surface and atmosphere. The influence of magnetic concentrations on the local temperature structure of the solar surface and atmosphere varies strongly with the size of the magnetic feature, as described below.

\begin{figure}
\includegraphics[width=\textwidth]{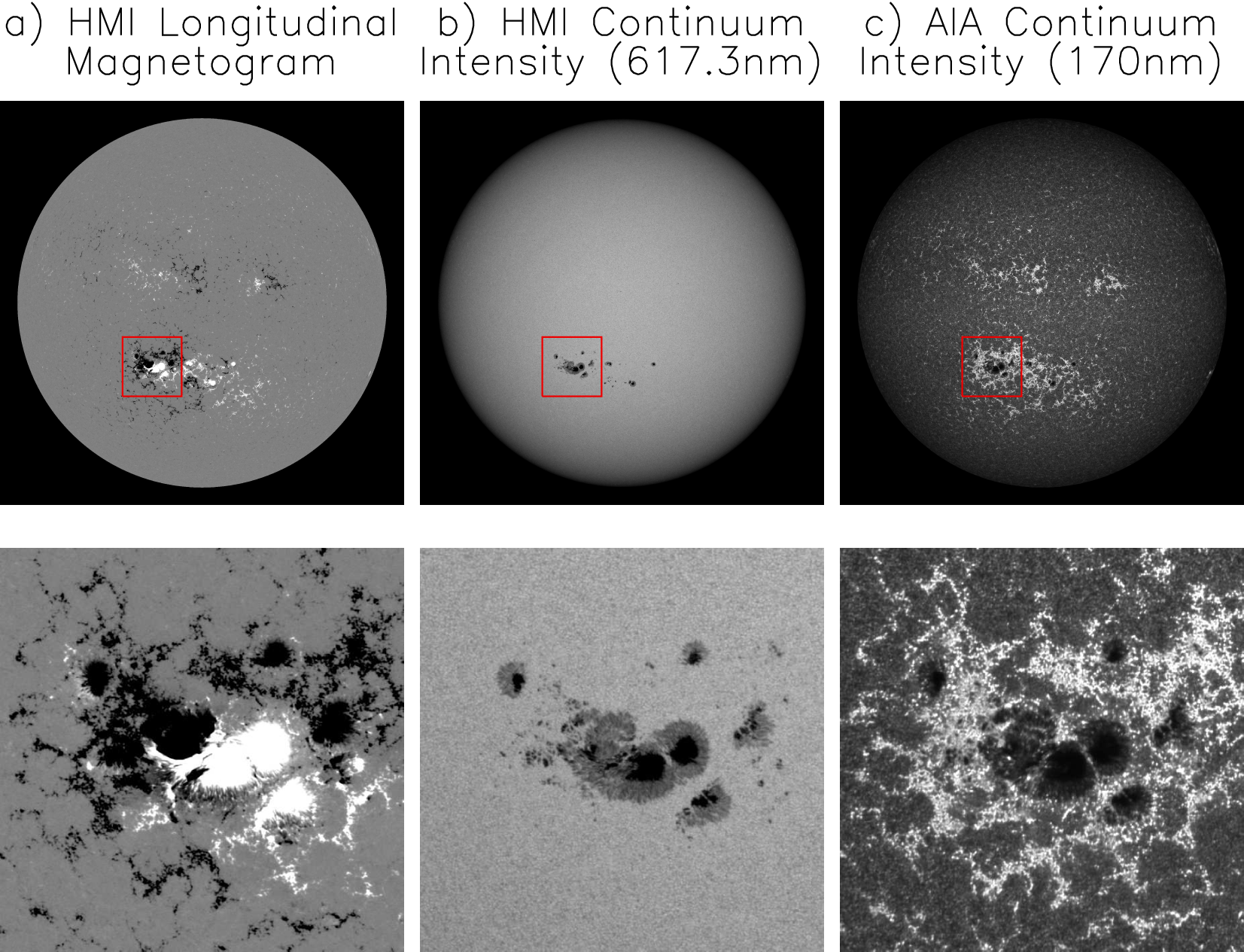}
\caption{Top: Concurrent observations, from 11th July 2012, of a) the line-of-sight magnetic field and the continuum intensity at b) 617.3 nm and c) 170 nm from the Helioseismic and Magnetic Imager, HMI \citep{schou12} and the Atmospheric Imaging Assembly, AIA \citep{lemen12} onboard the Solar Dynamics Observatory, SDO. The AIA image has been resampled to register with the HMI images. Bottom: Blow up of the boxed region, featuring active region NOAA 11520. From a) to c), the grey scale is saturated at $\pm30\:{\rm G}$, about $60\%$ and $120\%$, and $20\%$ and $300\%$ of the mean quiet Sun level at disc centre.}
\label{cma17}
\end{figure}

As a consequence of pressure balance, the interior of magnetic concentrations is evacuated. The lower density creates a depression in the optical depth unity surface and magnetic buoyancy, the result of which is flux tubes are largely vertical. The intensity contrast in the continuum is influenced by the competing effects of magnetic suppression of convection and radiative heating from surrounding granulation through the side walls of the depression. Sunspots and pores are dark from the magnetic suppression of convection within these features \citep[see Fig. \ref{cma17} and the reviews by][]{solanki03,rempel11}. For small-scale magnetic concentrations, this is overcome by the lateral heating, rendering them bright \citep{spruit76,spruit81,grossmann94,vogler05}, especially away from the disc centre as the side walls come into greater view \citep{spruit76,keller04,carlsson04,steiner05}. This feature gives rise to the bright faculae visible near the limb in white-light. The upper layers of the atmosphere enclosed within flux tubes are heated by mechanical and resistive dissipations \citep{musielak03,moll12}, and radiation from deeper layers \citep{knoelker91}. This enhances their intensity within spectral lines and in the ultraviolet, which are formed at greater heights than the visible continuum \citep[see Fig. \ref{cma17}c and][]{frazier71,mitchell91,morrill01,ermolli07,yeo13,rietmuller10}. A schematic of thin flux tubes and the mechanisms described above can be found in Fig. 5 of \cite{solanki13}.

Models describing the variation in solar irradiance at timescales greater than a day by the intensity deficit and excess facilitated by photospheric magnetism have achieved substantial success in reproducing measured solar irradiance \citep[see Sects. \ref{models2} and \ref{models3}, and][]{domingo09}. While other plausible mechanisms have also been proposed \citep{wolff87,kuhn88,cossette13}, related to physical processes in the solar interior, supporting evidence is not straightforward to obtain and consequently still largely lacking.

\begin{figure}
\includegraphics[width=\textwidth]{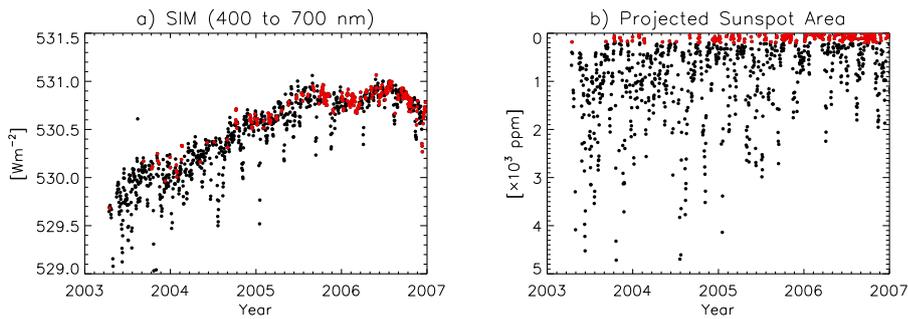}
\caption{a) Integrated solar irradiance in the visible in SIM SSI and b) the concurrent projected sunspot area from the composite record by \citealt{balmaceda09} (version 0613). Points corresponding to days where the projected sunspot area is less than 200 ppm are highlighted in red. The vertical axis of the projected sunspot area plot is inverted to aid interpretation.}
\label{comparesimsa}
\end{figure}

The increase in visible solar irradiance registered by SIM between 2003 and 2006 (Fig. \ref{comparesatsimssr}b) came at a time when solar activity was declining. For the solar cycle variation in photospheric magnetism to be compatible with this trend in visible solar irradiance, small-scale magnetic concentrations would have to be dark in the visible. The intensity deficit from sunspots and pores, while weakening over this period, is not driving the upward trend apparent in SIM visible solar irradiance. This is demonstrated in Fig. \ref{comparesimsa}a by the variation in SIM visible flux over the days with minimal sunspot activity, highlighted in red. If the increase in SIM visible flux is indeed from sunspot darkening, then it should be absent or at least markedly weaker in the days with minimal sunspot activity. As depicted in the figure, the overall trend in the SIM visible flux time series is no less manifest in the days where the projected sunspot area is minimal, below 200 ppm.

Turning now to small-scale magnetic concentrations, various studies have noted negative contrasts in the visible continuum near disc centre at both low and high magnetogram signal levels \citep[summarized in Table 2 of][]{yeo13}.

However, the results of recent investigations suggest that these apparent negative contrasts are associated with the fact that small-scale magnetic elements congregate mainly within dark intergranular lanes \citep{schnerr11,kobel11} and observational effects related to the limited spatial resolutions and telescope diffraction \citep{rohrbein11} than any indication that small-scale magnetic concentrations are dark in the visible continuum near disc centre.

Taken together with the intensity enhancement towards the limb and within spectral lines \citep{topka97,ortiz02,yeo13}, it is highly unlikely that small-scale magnetic concentrations, at least overall, might be dark in the visible. The upward trend in SIM visible solar irradiance between 2003 and 2006 is not consistent with our current understanding of solar surface magnetism and its effect on solar irradiance.

\subsection{Model architectures}
\label{models2}

There are two broad categories of solar irradiance models, distinguished by the modelling approach, commonly referred to as `proxy' and `semi-empirical'.

\subsubsection{Proxy models}
\label{models2a}

As stated in the introduction, the measurement of solar irradiance from space quickly revealed an apparent connection between TSI and the passage of active regions across the solar disc. This was followed by the development of models aimed at reconstructing solar irradiance by the multivariate regression of indices of solar activity to measured solar irradiance. The index data serve as proxies of the effects of bright and dark magnetic structures on the radiative output of the Sun, therefore the term proxy models.

Sunspot darkening is usually represented by sunspot area or the photometric sunspot index, PSI \citep{hudson82,frohlich94} and facular brightening by chromospheric indices such as the Ca II K \citep{keil98}, Mg II \citep{heath86} or F10.7 \citep[10.7 cm radio flux,][]{tapping87,tapping13} indices. In this context the term sunspot includes pores and the term faculae encompasses quiet Sun network. The Ca II K and Mg II indices are given by the ratio of the disc-integrated flux in the line core of the Ca II K line and the Mg II h and k doublet to that at nearby reference wavelengths. The line core to `continuum' ratio is preferred over absolute fluxes as it is more robust to instrument degradation.

The reconstructions of TSI by the group at the San Fernando Observatory, SFO \citep{chapman96,chapman12,chapman13,preminger02} employ sunspot and faculae indices derived from full-disc photometric images obtained at the observatory. These models proved to be particularly successful among proxy models. By employing full-disc imagery instead of Sun-as-a-star measures such as the indices listed above, they include the centre-to-limb variation of sunspot and faculae contrast, albeit only at the photometric bandpass. The latest iteration, based on visible red and Ca II K observations, reproduced most of the variation in TIM radiometry \citep[$R^2=0.95$,][]{chapman13}.

Since the proxy model approach relies on reliable solar irradiance measurements, it is not straightforward to reconstruct SSI by this method due to the long-term uncertainty of available measurements and the relative paucity of observations outside the ultraviolet (Sect. \ref{ssimeasurements}). Certain proxy models make use of the fact that the effects of instrument degradation on apparent variability is relatively benign at shorter timescales to circumvent long-term stability issues. The idea is to fit index data to measured rotational variability, either by detrending the index and SSI data or confining the regression to rotational periods, and then assume the indices-to-irradiance relationships so derived to all timescales \citep{lean97,pagaran09,thuillier12}. As we will discussed in Sect. \ref{discussionproxy}, the assumption that the underlying relationship between indices of solar activity and solar irradiance is similar at all timescales is not likely valid.

\subsubsection{Semi-empirical models}
\label{models2b}

The next level of sophistication in the modelling of solar irradiance is realized by semi-empirical models. In these models, the solar disc is segmented by surface feature type, termed `components'. The filling factor (proportion of the solar disc or a given area covered) and time evolution of each component is deduced from indices of solar activity or suitable full-disc observations. This information is converted to solar irradiance employing the intensity spectra of the various components. These are calculated applying spectral synthesis codes to semi-empirical model atmospheres of said feature types \citep{fontenla99,fontenla09,unruh99,shapiro10}. The reconstruction of the solar spectrum is given by the filling factor-weighted sum of the component intensity spectra. The semi-empirical model atmospheres describe the temperature and density stratification of the solar atmosphere within each component, constrained and validated by observations (therefore the term `semi-empirical').

The semi-empirical approach has the advantage that it yields SSI independent of the availability of reliable measurements. Additionally, for the models that rely on full-disc observations for the filling factor of the solar surface components (i.e., when the exact disc position of magnetic features is known), the centre-to-limb variation of the radiant behaviour of each component can be taken into account by generating and applying the corresponding intensity spectra at varying heliocentric angles.

\subsection{Present-day models}
\label{models3}

At present, there are five models aimed at reconstructing both TSI and SSI reported in the literature, reviewed in \cite{ermolli13}. These models describe the solar spectrum over at least from the far-ultraviolet to the infrared, such that the bolometric value is a close approximation of TSI. They are,
\begin{itemize}
	\item NRLSSI \citep[Naval Research Laboratory Solar Spectral Irradiance,][]{lean97,lean00},
	\item SATIRE-S \citep[Spectral And Total Irradiance REconstruction for the Satellite era,][]{unruh99,fligge00,krivova03,krivova06,krivova11a,yeo14b},
	\item SRPM \citep[Solar Radiation Physical Modelling,][]{fontenla99,fontenla04,fontenla06,fontenla09,fontenla11},
	\item OAR \citep[Observatorio Astronomico di Roma,][]{ermolli03,ermolli11,ermolli13,penza03} and
	\item COSI \citep[COde for Solar Irradiance,][]{haberreiter08,shapiro10,shapiro11,shapiro13}.
\end{itemize}
Apart from the NRLSSI, these models adopt the semi-empirical approach. In the following, we discuss the recent results obtained with the SATIRE-S model by \citealt{yeo14b} (Sect. \ref{modelsatires}) before giving an overview of the other models listed (Sect. \ref{modelsrpm}). For a broader review of models of solar irradiance, we refer the reader to \cite{domingo09,ermolli13,solanki13}.
 
\subsubsection{SATIRE-S}
\label{modelsatires}

The SATIRE-S model \citep{fligge00,krivova03,krivova11a} relies on spatially-resolved full-disc observations of magnetic field and intensity to segment the solar disc into quiet Sun, faculae, sunspot umbra and sunspot penumbra. It has been applied to longitudinal magnetograms and continuum intensity images collected at the KPVT \citep[in operation from 1974 to 2003,][]{livingston76,jones92}, as well as from SoHO/MDI \citep[1996 to 2011,][]{scherrer95} and SDO/HMI\footnote{In full, the Kitt Peak Vacuum Telescope (KPVT), the Michelson Doppler Imager onboard the Solar and Heliospheric Observatory (SoHO/MDI)), and the Helioseismic and Magnetic Imager onboard the Solar Dynamics Observatory (SDO/HMI).} \citep[2010 to the present,][]{schou12} to reconstruct total and spectral solar irradiance over various periods between 1974 and 2013 \citep{krivova03,wenzler06,ball12,ball14,yeo14b}. In the latest iteration \citep{yeo14b}, KPVT, MDI and HMI magnetograms were cross-calibrated in such a way that the model input from all the data sets combine to yield a single consistent TSI/SSI time series covering the entire period of 1974 to 2013 as the output. Apart from the NRLSSI, which extends back to 1950, this is the only other daily reconstruction of the solar spectrum spanning the ultraviolet to the infrared from present-day models to extend over multiple solar cycles.

At present, the model employs the intensity spectra of quiet Sun, faculae, umbra and penumbra from \cite{unruh99}, which were generated with the ATLAS9 spectral synthesis code \citep{kurucz93}. As ATLAS9 assumes local thermodynamic equilibrium (LTE), it fails in the ultraviolet below approximately $300\:{\rm nm}$. Ultraviolet solar radiation is formed in the upper photosphere and lower chromosphere, where the plasma is increasingly less collisional. The breakdown below 300 nm is accounted for by rescaling the 115 to 180 nm segment of the reconstructed spectra to SORCE/SOLSTICE SSI\footnote{This is relatively unaffected by the long-term stability issues plaguing available SSI measurements discussed in Sect. \ref{ssimeasurements1} as they only emerge towards longer wavelengths (Fig. \ref{satireuarssorcessr}).} and offsetting the 180 to 300 nm segment to the Whole Heliospheric Interval (WHI) reference solar spectra by \cite{woods09}, as detailed in \cite{yeo14b}. SATIRE-S is the only semi-empirical model to include the influence of departures from LTE through such a data-driven approach. The SRPM, OAR and COSI models take a more direct approach, making use of various non-LTE spectral synthesis codes to generate the intensity spectra of solar surface components. The various non-LTE codes differ from one another by the method non-LTE effects are approximated \citep[see][]{fontenla99,uitenbroek02,shapiro10}. 

Recall, due to uncertainties in the amplitude of solar cycle variation, the three published TSI composites exhibit conflicting decadal trends (Sect. \ref{tsimeasurements}). The TSI reconstructed by SATIRE-S is a significantly closer match to the PMOD composite than to the ACRIM or the IRMB composite, replicating most of the variability ($R^2=0.92$) over the entire length of the composites (Fig. \ref{comparetsi}b). Reconstructed TSI also exhibits excellent agreement with the measurements from individual instruments such as ACRIM3, TIM and VIRGO. The record from the PMO6V\footnote{VIRGO TSI is actually given by the combination of the measurements from two onboard radiometers, DIARAD and PMO6V. The solar cycle variation in the DIARAD and PMO6V records is nearly identical but they do differ at rotational timescales resulting in very different $R^2$ values on comparison with other measurements or models.} radiometer on VIRGO is particularly well-matched ($R^2=0.96$). The secular decline between the 1996 and 2008 solar cycle minima in VIRGO radiometry is reproduced to within $0.05\:\wms$ (Fig. \ref{comparetsi}a). This agreement between SATIRE-S TSI and VIRGO radiometry, which extends 1996 to the present, encompassing all of solar cycle 23, is significant. It implies that at least $96\%$ of the variability in solar irradiance over this period, including the secular variation between the 1996 and 2008 solar cycle minima can be explained by solar surface magnetism alone.

\begin{figure}
\includegraphics[width=\textwidth]{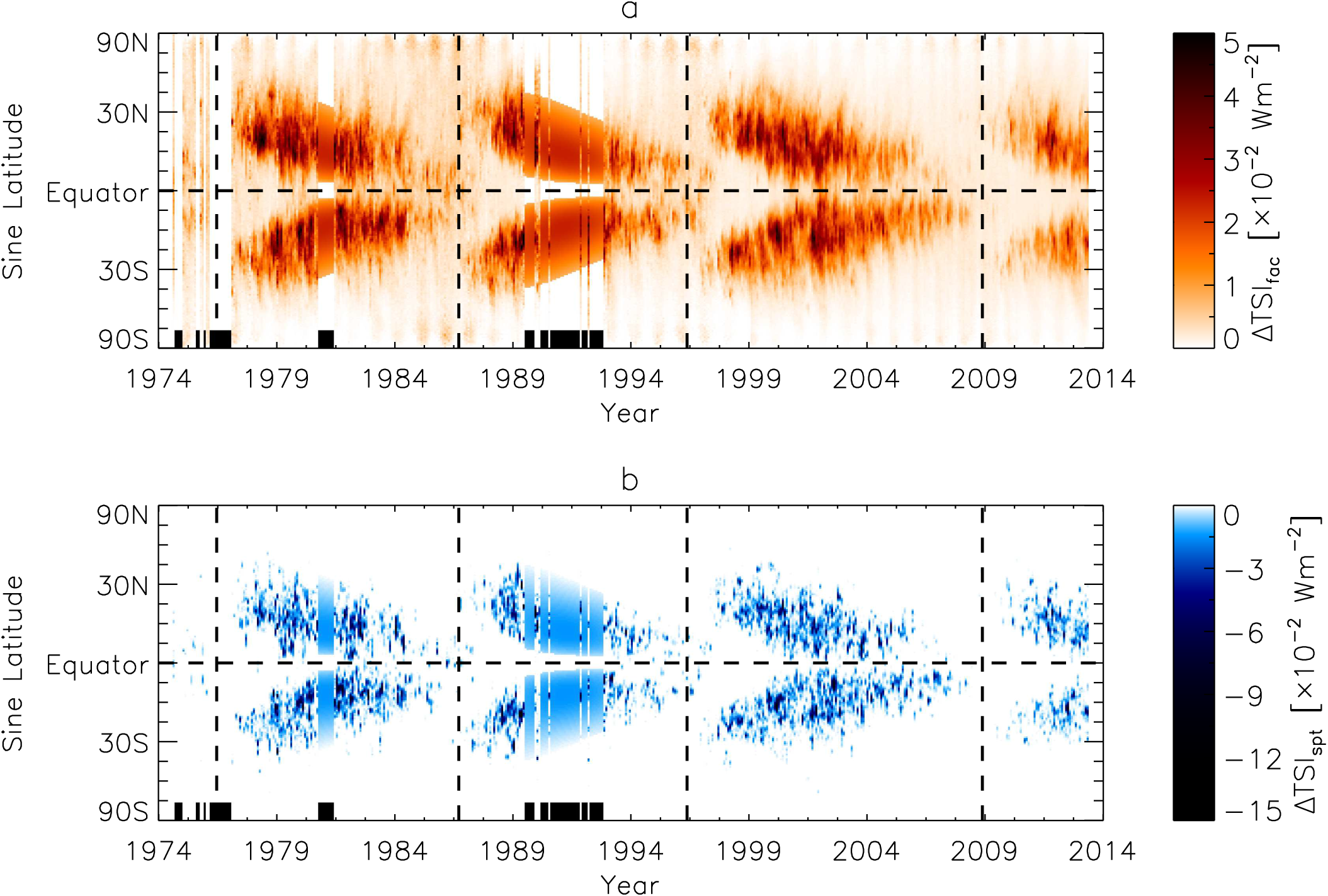}
\caption{Variation in TSI from a) faculae brightening, $\dtsifac$ and b) sunspot darkening, $\dtsispt$ in the SATIRE-S model, as a function of time and latitude (the monthly average in sine latitude intervals of 0.01). The black bars along the horizontal axes mark the months with no values from the lack of suitable magnetogram data. The gaps around the maxima of solar cycles 21 and 22 are filled by interpolation. The horizontal and vertical dashed lines denote the equator and epoch of solar cycle minima, respectively.}
\label{cpsbutterflyssr}
\end{figure}

The bolometric facular brightening and sunspot darkening, $\dtsifac$ and $\dtsispt$ with respect to the TSI level of the magnetically quiet Sun, binned and averaged by month and sine latitude, is expressed in Fig. \ref{cpsbutterflyssr}. This is an update of the similar figure by \cite{wenzler05} based on an earlier SATIRE-S reconstruction that employed KPVT data alone. Since solar surface magnetism is concentrated in active regions, it follows then that the latitudinal distribution of the associated intensity excess/deficit demonstrate Sp{\"o}rer's law, resembling butterfly diagrams of sunspot area/position and magnetic flux \citep[see e.g., Figs. 4 and 14 in][]{hathaway10}. A diagram similar to Fig. \ref{cpsbutterflyssr}b, based on the PSI, was presented by \citealt{frohlich13} (Fig. 5 in his paper).

\begin{figure}
\includegraphics[width=\textwidth]{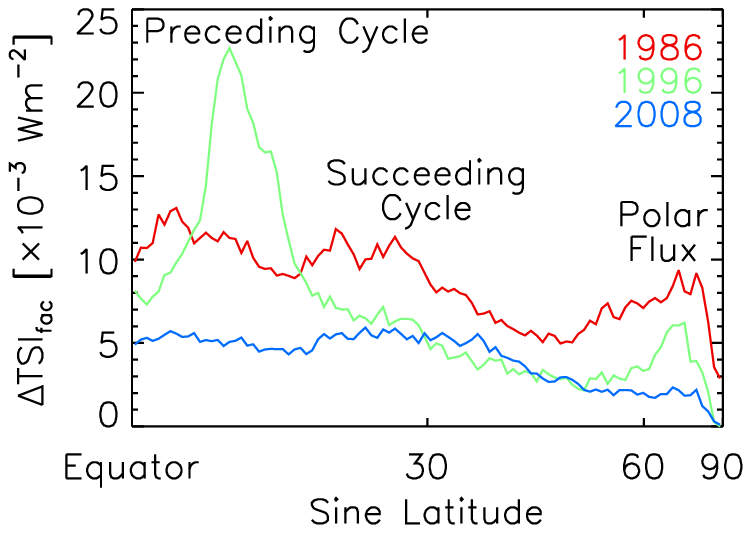}
\caption{Latitudinal distribution of facular brightening, $\dtsifac$ in SATIRE-S at the 1986, 1996 and 2008 solar cycle minima, taken from Fig. \ref{cpsbutterflyssr}a. Facular brightening is influenced by active regions in the low and mid-latitudes, associated with the preceding and succeeding solar cycles, respectively and polar flux at high latitudes.}
\label{secular}
\end{figure}

Since sunspots are largely absent around solar cycle minima, the minimum-to-minimum variation in SATIRE-S TSI is dominantly from the change in facular brightening. In Fig. \ref{secular}, we plot the latitudinal distribution of facular brightening at the last three solar cycle minima, taken from the butterfly diagram (Fig. \ref{cpsbutterflyssr}a). Around the 1986 solar cycle minimum, facular brightening is elevated close to the equator, around mid-latitudes and towards high latitudes (red curve). The broad peaks near the equator and at mid-latitudes correspond to active regions associated with the preceding cycle and the succeeding cycle, respectively. The increase towards high latitudes relates to magnetic elements transported polewards by meridional circulation over the course of the previous cycle (i.e., polar flux). The differing latitudinal distribution at the three solar cycle minima illustrates how the minimum-to-minimum trend in faculae brightening and therefore TSI is modulated by the prevailing magnetic activity in the three latitude regions. The low flat profile of the blue curve, corresponding to the 2008 solar cycle minimum, reflects the near-complete absence of any form of activity during this period, which contributed significantly to the secular decline between the 1996 and 2008 solar cycle minima.

\begin{figure}
\includegraphics[width=\textwidth]{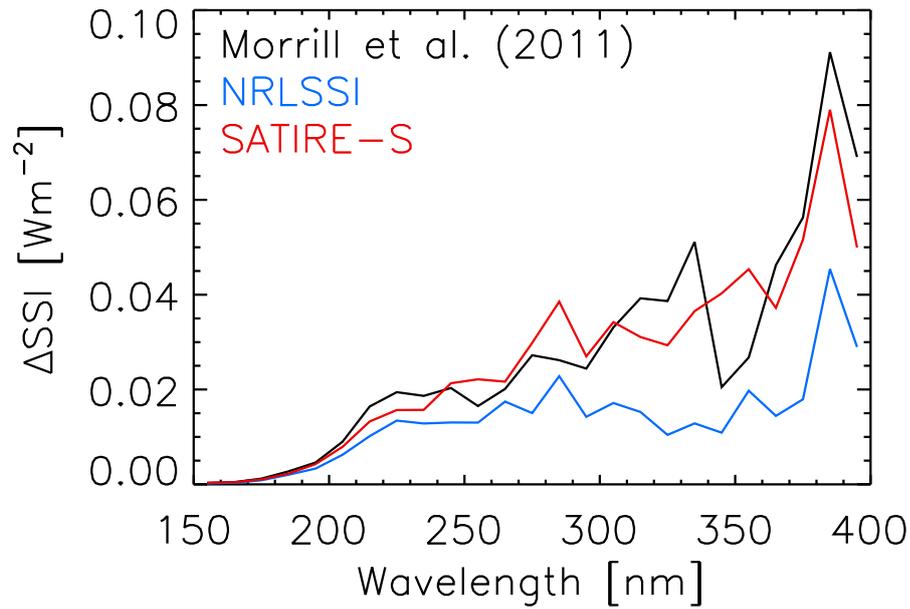}
\caption{The change in solar irradiance, ${\rm \Delta{}SSI}$, integrated over 10 nm intervals, between solar cycle maximum and minimum (as in Fig. \ref{barchart}), in the \cite{morrill11}, NRLSSI and SATIRE-S models.}
\label{comparesatmorrillssr}
\end{figure}

The ultraviolet solar irradiance measurements from the various instruments sent into space diverge in terms of the amplitude of solar cycle variation, especially above 240 nm and between SORCE and pre-SORCE missions (Sect. \ref{ssimeasurements1}). Consequently, while the SATIRE-S reconstruction reproduces the cyclical variability in the ultraviolet solar irradiance observations from the UARS and SORCE missions closely below 180 nm, above this wavelength, it replicates certain records better than others (Fig. \ref{satireuarssorcessr}). The reconstruction is a close match to the empirical model by \cite{morrill11}. This model, based on the matching of the Mg II index to SUSIM SSI, represents an estimation of SUSIM-like spectrometry with the stability corrected to that of the Mg II index. Notably, the amplitude of solar cycle variation is similar, even above 240 nm (Figs. \ref{barchart} and \ref{comparesatmorrillssr}). The reconstruction does not replicate the overall trend in SIM SSI (Fig. \ref{comparesatsimssr}). It is worth noting here that there is no model reported thus far that is able to reproduce the solar cycle variation in SIM SSI and TSI simultaneously.

\begin{figure}
\includegraphics[width=\textwidth]{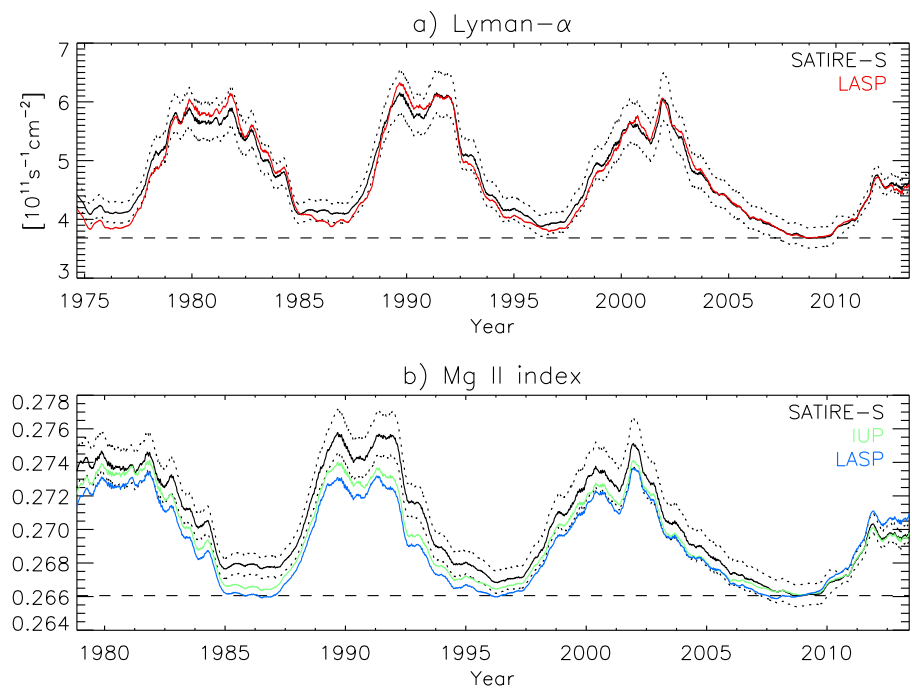}
\caption{a) The Lyman-$\alpha$ irradiance and b) Mg II index based on the SATIRE-S reconstruction of SSI (black solid lines). The reconstruction uncertainty is denoted by the dotted lines. Also illustrated, the LASP Lyman-$\alpha$ composite (red), and the IUP (green) and LASP (blue) Mg II index composites, regressed to the rotational variability and offset to the 2008 solar cycle minimum level (dashed lines) in the respective SATIRE-S time series. All the time series were smoothed with a 181-day boxcar filter. Adapted from \cite{yeo14b}.}
\label{comparesatssissr}
\end{figure}

The Lyman-$\alpha$ irradiance of the reconstruction reproduces most of the variability ($R^2=0.94$), including the solar cycle trend (Fig. \ref{comparesatssissr}a), in the LASP\footnote{The Laboratory for Atmospheric and Space Physics.} Lyman-$\alpha$ composite \citep{woods00}. The Mg II index derived from the reconstruction is highly correlated to the competing Mg II index composites by LASP \citep{viereck04,snow05b} and by IUP\footnote{IUP is the germanophone acronym of the Institute of Environmental Physics at the University of Bremen.} \citep{viereck99,skupin05c,skupin05b}, in particular with the latter ($R^2=0.96$). It was, however, less successful in replicating the decadal trend in these two composites (Fig. \ref{comparesatssissr}b). That said, it did reproduce, to within model uncertainty, the secular decline between the 1996 and 2008 solar cycle minima in the IUP Mg II index composite.

\subsubsection{Other present-day models}
\label{modelsrpm}

The present-day models capable of returning the solar spectrum over the ultraviolet to the infrared are, with the exception of the SRPM, broadly consistent with one another. They differ in certain aspects, most notably in terms of the spectral dependence of the cyclical variability (Fig. \ref{barchart}). In this section, we will give a brief description of the NRLSSI, SRPM, OAR and COSI models, and highlight the key discrepancies between these models and SATIRE-S.

The NRLSSI \citep{lean97,lean00} describes the effect of sunspot darkening and faculae brightening on a model spectrum of the quiet Sun. The time evolution of sunspot darkening is given by the PSI derived from sunspot area records and facular brightening by the Ca II K, Mg II and F10.7 indices. In the ultraviolet (120 to 400 nm), the variation in solar irradiance is inferred from the multivariate regression of the index data to the rotational variability in UARS/SOLSTICE SSI. This is achieved by detrending both index and SSI data prior to the regression. Above 400 nm, it is given by the sunspot and faculae contrast models from \cite{solanki98a}, modulated in time by the index data.

The models discussed here, apart from the SRPM, all see reconstructed SSI varying in phase with the solar cycle in the ultraviolet and visible (Fig. \ref{barchart}). In the infrared, facular contrast is weak and negative at certain wavelengths, depending on the model. This allows sunspot darkening to dominate such that the overall level at activity maximum can be lower than at minimum as illustrated for SATIRE-S in Fig. \ref{comparesatsimssr}d. Depending on the sunspot/facular contrast adopted by the various models, they differ in the wavelength range and strength of this effect \citep[see Fig. 7 in][]{ermolli13}. This effect is relatively weak in the NRLSSI such that the integrated flux over the shortwave-infrared (1000 to 2430 nm) still varies in phase with the solar cycle, contrary to the other models (Fig. \ref{barchart}). The variation over the solar cycle between 240 and 400 nm is also weaker, almost half of that in SATIRE-S, attributed to confining the regression to rotational variability \citep{ermolli13}. The consistency between SATIRE-S and the SUSIM-based model of \cite{morrill11} gives further support to the amplitude of solar cycle variation in the ultraviolet exhibited by these two models (Fig. \ref{comparesatmorrillssr}). While SATIRE-S replicates the secular decline between the 1996 and 2008 solar cycle minima in VIRGO TSI radiometry (Fig. \ref{comparetsi}a), the NRLSSI does not as this minimum-to-minimum variation is absent in the Mg II index composite employed, that released by LASP (Fig. \ref{comparesatssissr}b).

The SRPM denotes the set of data and tools for semi-empirical modeling of solar irradiance, including a non-LTE spectral synthesis code, developed by \cite{fontenla99,fontenla04,fontenla06,fontenla09,fontenla11}. Currently, the package features semi-empirical model atmospheres for nine solar surface components: quiet Sun internetwork, quiet Sun network lane, enhanced network, plage, bright plage, sunspot umbra and sunspot penumbra, presented in \cite{fontenla09}, and dark quiet Sun internetwork and hot facula, introduced later in \cite{fontenla11}. \cite{fontenla11} made various adjustments to the \citealt{fontenla09} model atmospheres guided by SORCE SSI. The solar irradiance reconstruction presented by the authors, extending 2000 to 2009, is based on these modified model atmospheres. The filling factors of the various components were derived from full-disc images in the red part of the visible spectrum and Ca II K acquired with the Precision Solar Photometric Telescope, PSPT at OAR \citep{coulter94,ermolli98}.

The modifications to the \cite{fontenla09} model atmospheres resulted in the reconstructed visible solar irradiance registering lower levels during periods of higher activity \citep[Figs. 6 and 9 in][]{fontenla11}, qualitatively consistent with the increase in SIM visible flux between 2003 and 2006 (Fig. \ref{comparesatsimssr}b). However, the solar cycle variation in the ultraviolet is still weaker than what was recorded by SORCE/SOLSTICE. The model thus failed to reproduce the solar cycle variation in measured TSI, which all the other present-day models are able to do with reasonable success. These shortcomings were taken by the authors to imply that the number of solar surface components is insufficient, which runs counter to the OAR results.

The team at OAR also employed the full-disc images recorded with the PSPT at the observatory in the series of proxy and semi-empirical reconstructions of solar irradiance reported \citep{penza03,domingo09,ermolli11}. Work on a new semi-empirical model is in progress \citep{ermolli13}. This latest effort considers the seven solar surface components defined in \cite{fontenla09}. The filling factors were obtained from PSPT observations covering the period of 1997 to 2012 following \cite{ermolli10}. The intensity spectra corresponding to each component were calculated with the \cite{fontenla09} model atmospheres, without the more recent modifications introduced by \cite{fontenla11}, using the non-LTE spectral synthesis code RH\footnote{Based on the work of and abbreviated after \cite{rybicki91,rybicki92}.} \citep{uitenbroek02}. The computation is therefore, apart from employing the \cite{fontenla09} model atmospheres as is, broadly analogous to the SRPM reconstruction presented by \citealt{fontenla11}.

In the ultraviolet and visible, the OAR reconstruction is roughly consistent with NRLSSI, SATIRE-S and COSI, exhibiting a similar disparity with SIM SSI (Fig. \ref{barchart}). Also in line with these other models, the TSI from the reconstruction replicates most of the variability in TIM radiometry and the PMOD composite. The agreement with measured TSI suggests that the seven solar surface components described in \cite{fontenla09} are sufficient for the semi-empirical modeling of at least TSI, in contradiction to the conclusion of \cite{fontenla11}. The OAR results also support the conclusions of \cite{ball11,deland12,lean12,unruh12,yeo14b}, that the discrepancy between SIM SSI and models arise from unaccounted instrumental effects in the SIM record and does not warrant a significant rethink in how solar irradiance is modelled.

The non-LTE spectral synthesis code COSI \citep{haberreiter08,shapiro10} has been utilised to generate intensity spectra of solar surface components for semi-empirical modeling of solar irradiance \citep{haberreiter05}. The current implementation utilizes intensity spectra generated with the model atmospheres by \cite{fontenla99}. These were applied to sunspot number, $^{10}{\rm Be}$ and neutron monitor data to reconstruct solar irradiance back to the Maunder minimum period and over the Holocene at 1-year and 22-year cadence, respectively \citep{shapiro11}. More recently, they were also applied to SRPM PSPT-based filling factors \citep{shapiro13}, and HMI full-disc longitudinal magnetograms and continuum intensity images \citep{thuillier14b}.

Contrary to the NRLSSI, SATIRE-S and OAR models, the \cite{shapiro11} reconstruction varied in anti-phase with the solar cycle in the near-infrared (700 to 1000 nm, Fig. \ref{barchart}), accompanied and compensated by an enhanced variability in the ultraviolet. This was attributed to the use of a single model atmosphere each for sunspots and for plage, so not distinguishing between sunspot umbra and penumbra and between plage and bright plage \citep{ermolli13}. To account for this simplification, \cite{shapiro13} reduced sunspot and plage contrast in such a manner that brought the reconstruction presented into alignment with SORCE/SOLSTICE measurements in the Herzberg continuum (190 to 222 nm). A more comprehensive approach is in development.

\section{Reconciling measurements and models}
\label{discussionssr}

\subsection{Proxy models}
\label{discussionproxy}

Solar irradiance observations, in particular SSI, suffer non-trivial long-term uncertainty (Sect. \ref{measurements}). As mentioned in Sect. \ref{models2a}, in the regression of indices of solar activity to measured SSI, certain proxy models confine the fitting to rotational variability to circumvent bias from instrumental trends. Such a step implicitly assumes that the relationship between the two is similar at all timescales.

While the rotational variability in solar irradiance is largely driven by the time evolution of active regions, the variability at longer timescales is dominated by the magnetic flux distributed in the quiet Sun \citep{foukal88,fligge00,solanki02}. The response of chromospheric indices to magnetic flux in active regions and in the quiet Sun is evidently not the same \citep{tapping87,solanki04,ermolli10,foukal11}. The weak solar cycle variation in the ultraviolet in the NRLSSI highlights the limitation of applying the relationship between chromospheric indices and solar irradiance at rotational timescales to longer timescales.

VIRGO registered a secular decline in TSI of over $0.2\:\wms$ between the 1996 and 2008 solar cycle minima, approximately $20\%$ of the solar cycle amplitude \citep{frohlich09}. The NRLSSI does not replicate this clear secular trend in VIRGO radiometry as it is absent in the LASP Mg II index composite. \cite{frohlich09,frohlich12,frohlich13} attributed the discrepant decadal trend in the LASP Mg II index composite and VIRGO TSI to a possible cooling/dimming of the photosphere between the two solar cycle minima. This disparity is more likely related to the non-linear relationship between chromospheric indices and solar irradiance, discussed in the previous paragraph \citep{foukal11} and the long-term uncertainty of Mg II index data. While the LASP Mg II index composite is effectively level between the 1996 and 2008 solar cycle minima, the competing composite by IUP does exhibit a secular decline (Fig. \ref{comparesatssissr}b). This minimum-to-minimum drop in the IUP composite is qualitatively replicated in the Mg II index produced from the SATIRE-S reconstruction, the TSI from which reproduces the secular decline in VIRGO TSI (Sect. \ref{modelsatires}). The discrepancy between the IUP and LASP composites demonstrates how even for an activity proxy as robust to instrumental effects as the Mg II index, the long-term uncertainty can still be sufficient to obscure the underlying decadal variation.

The rigorous reconstruction of solar irradiance through proxy models would require a greater understanding of the relationship between indices of solar activity and solar irradiance, and of the long-term stability of index data, which is still largely unknown.

\subsection{Semi-empirical models}

Present-day semi-empirical models, reviewed in Sect. \ref{models3}, all employ one-dimensional or plane-parallel model atmospheres. Various studies have pointed out that the intensity spectra synthesized from one-dimensional representations of the spatially inhomogeneous solar atmosphere do not necessarily reflect the true average property \citep{uitenbroek11,holzreuter13}.

The intensity contrast of network and facular magnetic features varies with distance from disc centre and magnetic flux. In SATIRE-S, the magnetic flux dependence is linear with the magnetogram signal up to a saturation level, the free parameter in the model \citep{fligge00}. For the SRPM, OAR and COSI models, which employ full-disc Ca II K images, after identifying sunspots, the rest of the solar disc is segmented by the Ca II K intensity into multiple components. These measures are not only empirical but also do not properly account for the observation that the continuum and line core intensity contrast of small-scale magnetic concentrations scale differently with magnetogram signal \citep{yeo13}. As set out by \cite{unruh09}, three-dimensional model atmospheres would allow the possibility to relate the appropriate calculated intensity spectra to the magnetogram signal or Ca II K intensity directly.

In the continuum, the intensity contrast of small-scale magnetic elements increases with distance from disc centre before declining again close to the limb while the converse is observed within spectral lines. This difference comes primarily from the differing interaction between the line-of-sight, and magnetic flux tubes and the intervening atmosphere at the continuum and spectral lines formation heights \citep{solanki98b,yeo13}. In employing one-dimensional model atmospheres, present semi-empirical models do not capture these effects.

Three-dimensional model atmospheres based on observations \citep{socasnavarro11} and magnetohydrodynamics (MHD) simulations \citep{vogler05}, while growing in sophistication and realism, cannot as yet reproduce observations at all heights \citep{afram11}. A limiting factor is our understanding of the effects of spatial resolution, that is, the point spread function and how it is sampled by the imaging array, on observations \citep{danilovic08,danilovic13,rohrbein11}. This is especially severe for the small-scale magnetic concentrations which make up network and faculae as they are largely unresolved in current observations. The increasing availability of atmospheric seeing-free observations from space and balloon-borne missions, in particular high spatial resolution imagery, such as those from SUNRISE \citep{solanki10,barthol11}, will provide stringent constraints on model atmospheres. Space and balloon-borne telescopes have the advantage that the point spread function can be well-constrained \citep{mathew07,mathew09,wedemeyerbohm08,yeo14a}, rendering them particularly useful for this purpose.

Another source of uncertainty is the treatment of non-LTE effects, which are highly complex and not fully understood. The SRPM, OAR and COSI models employ various non-LTE spectral synthesis codes which differ by the approach taken to approximate non-LTE effects. In SATIRE-S, which relies on an LTE spectral synthesis code, non-LTE effects are accounted for empirically instead. While inexact, the one-dimensional model atmospheres and the various measures taken to account for non-LTE effects, direct or empirical, in present-day semi-empirical models are a practical necessity. As these simplifications are tested against and so constrained by observations, current models are found to be reasonable for the purpose of solar irradiance reconstruction. This is demonstrated by the broad consistency between reconstructed solar irradiance and measurements (Sect. \ref{models3}).

\section{Summary}
\label{summaryssr}

The TSI observations from the succession of radiometers sent into space since 1978 readily reveal solar cycle modulation. The clear detection of such modulations, only about $0.1\%$ of the overall level, is a remarkable achievement. The records from the various instruments do, however, differ in terms of the absolute level and the apparent amplitude of solar cycle variation. This is chiefly from the difficulty in accounting for instrument degradation. While the absolute radiometry of present-day instruments is converging due to the collaborative efforts of various teams, significant uncertainty persists over the long-term stability. This is evident in the conflicting decadal trends exhibited by the three published TSI composites.

Like TSI, ultraviolet (120 to 400 nm) solar irradiance has been monitored from space, almost without interruption, since 1978. Spectrometry is obviously a more complicated measurement. Not surprisingly, the uncertainty in the absolute radiometry and the amplitude of solar cycle variation is more severe than with TSI. Compounded by the wavelength dependence of instrumental influences, this translates into uncertainty in the spectral dependence of the cyclical variability. The problem is particularly acute above 240 nm and between measurements from the SORCE satellite and preceding missions.

The SIM instrument onboard SORCE provides what is still the only extended (2003 to 2011) record of SSI spanning the ultraviolet to the infrared (240 to 2416 nm) available. The measurements from the first few years of operation (2003 to 2008) saw ultraviolet solar irradiance declining almost twice as rapidly as TSI and visible solar irradiance ascending, in apparent anti-phase with the solar cycle. These trends conflict with projections from other measurements and models. Looking at the full period, the overall trend shows no obvious solar cycle modulation. The total flux recorded by the instrument, which surveys a wavelength range responsible for more than $97\%$ of the power in solar radiative flux, also fails to replicate the solar cycle variation evident in TSI.

Satellite monitoring of solar irradiance has been accompanied by the development of models aimed at recreating the observed variability. Solar irradiance is modulated by photospheric magnetism from its effect on the thermal structure and consequently the radiant behavior of the solar surface and atmosphere. Models of solar irradiance based on the assumption that variations at timescales greater than a day are driven by solar surface magnetism have achieved considerable success in replicating observations.

There are two broad categories of solar irradiance models, termed proxy and semi-empirical. Proxy models are based on the regression of indices of solar activity to solar irradiance observations. Semi-empirical models employ the intensity spectra of solar surface features calculated from semi-empirical model atmospheres with spectral synthesis codes. These intensity spectra are combined with the apparent surface coverage of the various features, derived from index data or full-disc observations, to reconstruct the solar spectrum.

We discussed the present-day models capable of returning the spectrum over the ultraviolet to the infrared: NRLSSI, SATIRE-S, SRPM, OAR and COSI. Apart from the NRLSSI, these models adopt the semi-empirical approach.

In the regression of index data to solar irradiance observations, certain proxy models including the ultraviolet segment of NRLSSI restrict the fitting to the rotational variability to factor out any bias from the long-term uncertainty of the solar irradiance measurements employed. In doing so, these models implicitly assume that the relationship between chromospheric indices, utilized in these models as a proxy of facular brightening, and solar irradiance at rotational timescales is applicable to longer timescales. Likely a consequence of the fact that the relationship between chromospheric indices and solar irradiance is really non-linear, the amplitude of solar cycle variation in the ultraviolet in the NRLSSI is weaker than in other present-day models.

Another limitation of the proxy approach is the fact that the reconstructed solar irradiance adopts the variability of the index records used in the reconstruction, along with the associated uncertainty. We argued that the long-term uncertainty of Mg II index data might be the reason why the NRLSSI does not replicate the secular decline between the 1996 and 2008 solar cycle minima in VIRGO TSI radiometry.

The SRPM reconstruction of solar irradiance presented by \cite{fontenla11} is the only one where visible flux varied in anti-phase with the solar cycle, in qualitative agreement with early SIM observations. This was achieved with modifications to the \cite{fontenla09} model atmospheres. However, the TSI from the reconstruction failed to replicate the solar cycle variation in measured TSI. The other models reviewed see visible solar irradiance varying in-phase with the solar cycle and reproduced TSI variability, including the solar cycle modulation, with reasonable success. Significantly, the OAR computation is, apart from the use of the \cite{fontenla09} model atmospheres without any modifications, largely analogous to the \cite{fontenla11} study in terms of the approach.

Considering the role of photospheric magnetism in driving variations in solar irradiance, the increase in the visible registered by SIM during its early operation, coming at a time where solar activity is declining, requires that small-scale magnetic concentrations be darker than the quiet Sun in this spectral region. However, our current understanding of the radiant properties of these solar surface features point to the converse.

Apart from the NRLSSI, the semi-empirical model SATIRE-S, recently updated by \cite{yeo14b}, gives the only other daily reconstruction of the solar spectrum spanning the ultraviolet to the infrared from present-day models to extend multiple solar cycles, covering 1974 to 2013. Of the three divergent TSI composites, the model found the greatest success in replicating the solar cycle variation in the PMOD composite. The TSI reconstruction is also a good match to present-day measurements, reproducing about $96\%$ of the variability in the PMO6V record and the secular decline between the 1996 and 2008 solar cycle minima in VIRGO radiometry. The SSI reconstruction replicates the solar cycle variation in the ultraviolet solar irradiance observations from the UARS and SORCE missions very closely below 180 nm. As the amplitude of solar cycle variation is poorly constrained in available SSI measurements at longer wavelengths, particularly above 240 nm, SATIRE-S, as with all other models, cannot exactly replicate SORCE solar cycle variation very well there. The amplitude of solar cycle variation in the reconstruction does however, match closely to the empirical model of \cite{morrill11}, which represents an approximation of SUSIM-like SSI with the stability corrected to that of the Mg II index. The model also replicates the solar cycle variation in the LASP Lyman-$\alpha$ composite and the secular decline between the 1996 and 2008 solar cycle minima in the IUP Mg II index composite.

The intensity spectra of solar surface features employed in present-day semi-empirical models are derived from one-dimensional model atmospheres which do not capture all the complexities of the radiant behaviour of the solar surface and atmosphere. Three-dimensional model atmospheres, though increasingly realistic, still cannot reproduce observations at all heights. Their development is impeded by the limited availability of high spatial resolution observations and the challenge in understanding instrumental influences on apparent radiance. Current semi-empirical models account for non-LTE effects either empirically by offsetting/rescaling reconstructed spectra to measured SSI or directly by employing non-LTE spectral synthesis codes, neither of which is exact. Constrained by observations, the intensity spectra of solar surface features generated from one-dimensional model atmospheres and present non-LTE schemes, are still somewhat reliable for the intended purpose. This is demonstrated by the broad consistency between the various semi-empirical models and their success in replicating measurements.

The direct observation of solar irradiance is a challenging endeavour. At present, the body of spaceborne measurements is still afflicted by uncertainties in the absolute radiometry, secular variation and spectral dependence of the cyclical variability. However, one cannot discount the considerable progress made over the past four decades with the collective effort of the community. A good example is the collaborative efforts which led to the convergence of ACRIM3, TIM and PREMOS absolute TSI radiometry. Models of solar irradiance based on solar surface magnetism have proved to be an able complement, augmenting our understanding of the observations and the physical processes underlying solar cycle variation in solar irradiance. While open questions remain, continual observational and modeling efforts will undoubtedly see the emergence of a more cohesive picture of solar cycle variation in solar irradiance.

\begin{acknowledgements}
We are thankful to Yvonne Unruh for her help in preparing Fig. \ref{barchart}, and Ilaria Ermolli and Karl-Heinz Glassmeier for the helpful discussions. The authors are also grateful to the International Space Science Institute for their invitation to and tremendous hospitality over the `The Solar Activity Cycle: Physical Causes And Consequences' workshop. Acknowledgment goes to the various individuals and teams responsible for the observations and models featured in this review. K.L.Y. is a postgraduate fellow of the International Max Planck Research School for Solar System Science. This work is partly supported by the German Federal Ministry of Education and Research under project 01LG1209A and the Ministry of Education of Korea through the BK21 plus program of the National Research Foundation, NRF.
\end{acknowledgements}

%\bibliographystyle{aps-nameyear}
%\bibliography{references}
%\nocite{*}

\end{document}